# Relaxor ferroelectric behavior and intrinsic magnetodielectric behavior near room temperature in $Li_2Ni_2Mo_3O_{12}$, a compound with distorted honeycomb and spin-chains

**Sanjay Kumar Upadhyay, Kartik K Iyer, Smita Gohil, Shankar Ghosh, P.L. Paulose, and E.V. Sampathkumaran***

Tata Institute of Fundamental Research, Homi Bhabha Road, Colaba, Mumbai 400005, India
(*corresponding author: sampathev@gmail.com)

Keeping current interests to identify materials with intrinsic magnetodielectric behaviour near room temperature and with novel pyroelectric current anomalies, we report temperature and magnetic-field dependent behavior of complex dielectric permittivity and pyroelectric current for an oxide, $Li_2Ni_2Mo_3O_{12}$, containing magnetic ions with (distorted) honey-comb and chain arrangement and ordering magnetically below 8 K. The dielectric data reveal the existence of relaxor ferroelectricity behaviour in the range 160 -240 K and there are corresponding Raman mode anomalies as well in this temperature range. Pyrocurrent behavior is also consistent with this interpretation, with the pyrocurrent peak-temperature interestingly correlating with the poling temperature. $^7Li$ NMR offer an evidence for crystallographic disorder intrinsic to this compound and we therefore conclude that such a disorder is apparently responsible for the randomness of local electric field leading to relaxor ferroelectric property. Another observation of emphasis is that there is a notable decrease in the dielectric constant with the application of magnetic field to the tune of about -2.4% at 300 K, with the magnitude varying marginally with temperature. Small loss factor values validate the intrinsic behaviour of the magnetodielectric effect at room temperature.

**(A version of this article will appear in print in Scientific Reports)**

The search for materials exhibiting large magnetodielectric (MDE) coupling in a broad temperature range, particularly around room temperature continues unabated in the literature. It is of great interest to identify such materials with this property at room temperature to facilitate applications [see, for instance, refs. 1-5]. It is now well-known that, around room temperature, many insulating materials become sufficiently conducting to result in detectable magnetoresistance. Under such circumstances, the debate of intrinsic versus extrinsic behavior of magneto-capacitance often arises[6]. In this article, keeping such trends of research activities in mind, we present the results of our investigation in this direction on a Li-based compound, $Li_2Ni_2Mo_3O_{12}$, which was not paid any attention in the literature after initial magnetic studies by Hase et al[7].

In the compound, $Li_2Ni_2Mo_3O_{12}$, Ni has been shown to possess magnetic moment (spin = 1), undergoing a complex magnetic order below $T_C$= 8 K [ref. 7]. This system consists of linear chains as well as distorted honey-comb of the divalent Ni ions. Thus, there are two independent sites for Ni ions. Through neutron diffraction measurements, it has been found that the Ni chains exhibit ferromagnetic order, while distorted honeycomb of Ni shows the antiferromagnetic structure. The nominal composition of the title compound considering disorder[8] can be written as $Li(Li_{0.5}Ni_{1.5})(Li_{0.5}Ni_{0.5})Mo_3O_{12}$. The crystal structure [orthorhombic, space group *Pmcn (No. 62)*] in which the compound forms is related to that of $NaCo_{2.31}Mo_3O_{12}$ [ref. 9]. The detailed crystal structure and magnetic structure of the title compound can be found in the references 8 and 7 respectively.

It is important to note that the magnetic properties of compounds with honey-comb arrangement [see, for instance, refs 10-14] of atoms and/or spin-chain systems [see, for instance, references 15-19] have been attracting a lot of attention experimentally as well as theoretically. However, there are only a few reports in the literature, addressing the dielectric and magneto-dielectric behavior of honey-comb and spin-chain-based oxides [see, for instance, refs 15-19]. Since the $Li_2Ni_2Mo_3O_{12}$ was reported to be an insulator, interestingly containing (distorted) honeycomb lattice and spin-chain features[7], we considered it worthwhile to explore its detailed dielectric and pyroelectric properties. With this motivation, we have carried out dielectric, magnetodielectric and pyrocurrent (*I*) studies as a function of temperature (*T*) on this compound. In order to supplement the arguments, we have also carried out magnetization (*M*), heat-capacity (*C*), Raman spectroscopy and $^7Li$ (nuclear spin = 3/2) NMR studies.

**Results**
**X-ray diffraction.** The x-ray diffraction (XRD) pattern is shown in Supplementary Information (see Supplementary Fig. S1). Within the detection limit (<2%) of the XRD measurement, the polycrystalline sample was found to be single phase (*Pmcn* space group). The experimental data was subjected to Rietveld refinement by using Fullprof programme[20] and the Wyckoff positions have been taken from ref. 8. The lattice parameters (*a*= 10.4118(4) Å, *b*= 17.4958(4) Å, *c*= 5.0778(4) Å) are in good agreement with the literature[8].

**Magnetization.** Since dc magnetization (*M*) behavior has been reported in ref. 7, we present here essential features from our data for the sake of completeness. We obtained *T*-dependence of dc magnetic susceptibility ($\chi$) in the presence of various magnetic fields (*H*= 100 Oe and 5 kOe). Our results are in good agreement with those reported [ref. 7]. The readers may see Supplementary Information for the $\chi(T)$ behavior obtained in 5 kOe (see Supplementary Fig. S2(a)) for the zero-field-cooled (ZFC) condition (from 300 K to 2 K) of the specimen. The curve was fitted with the Curie-Weiss law in the interval of 50 - 300 K. The value of the effective magnetic moment obtained from this fitting is about 3.23 $\mu_B$ per Ni atom, which is higher than that expected (2.83 $\mu_B$) for divalent Ni (S=1). We believe that such a marginally higher value could be the result of short-range magnetic correlations extending to such a high temperature range and intrinsic crystallographic disorder. The paramagnetic Curie-temperature, if inferred from this linear region, is about -40 K, and the negative sign indicates the dominance of antiferromagnetic correlations. These values agree with those reported in the literature[7]. Fig. 1a shows the ZFC and field-cooled (FC) $\chi$ curves obtained in 100 Oe. ZFC and FC curves exhibit a sudden upturn at 8 K with a lowering of temperature, due to the onset of magnetic ordering, followed by a peak around 7.2 K and finally a decrease. On further lowering of temperature, the sign of magnetization gets reversed (i.e. negative magnetization) around 3.6 K – the compensation temperature. ZFC and FC curves deviate only marginally from each other below 8 K, with a sign-crossover around the compensation temperature (3.6 K). From the absence of bifurcation of ZFC-FC curves, we conclude that this compound is not a spin-glass, despite the existence of crystallographic disorder. We carried out ac χ studies and there is a frequency (ν) independent peak in the real part (not shown here) at the onset of magnetic order, without any feature in the imaginary part; this establishes the absence of any kind of magnetic frustration.

We have also performed isothermal magnetization $M(H)$ studies up to 140 kOe (see Supplementary Fig. S2) at selected temperatures (1.8, 6, 10 and 20 K). At 1.8 K, the virgin curve lies outside the envelope curve and starts with the negative value (see top inset of Fig. S2(b)). This finding is agreeing with those reported for 1.7 K by Hase et al[7] who performed the studies up to 50 kOe. We would like to stress that an extension of isothermal $M$ studies to higher fields to 140 kOe in this investigation reveals another interesting feature - that is, a weak hysteresis, in the range 25 to 50 kOe in the curve for 1.8 K (see Fig. S2(c)). This finding implies the existence of disorder-broadened metamagnetic transition, which could be responsible for the virgin curve lying outside the envelope curve. Additionally, in the $M(H)$ curve at 6 K, a small hysteresis is observed in the low-field region (<2 kOe) (see Fig. S2(d)). Isothermal $M(H)$ plots continue to increase without any evidence for saturation at high fields in the magnetically ordered state. These findings, indicating the existence of both ferromagnetic and antiferromagnetic components, are consistent with the complex magnetic structure proposed in ref. 7.

**Heat capacity.** Fig. 1b shows the plot of $C(T)$ below 290 K and two features are worth noting - one broad feature around 225 K, and a sharper one near 8 K (see also top inset of Fig. 1b). These are seen in $C/T$ plots also (see bottom inset of Fig. 1b). The 8K-peak arises from long range magnetic ordering[7]. The high temperature peak, not reported earlier[7], may be correlated to relaxor ferroelectric behavior, reported in this article.

**Complex permittivity.** Complex permittivity studies have been performed from room temperature to 2 K. Dielectric constant ($\varepsilon'$) and the loss factor (tan$\delta$) as a function of temperature, obtained with various frequencies (10, 20, 30, 50, 70 and 100 kHz), are shown in Fig. 2a and Fig. 2b. The observation we would like to stress is that $\varepsilon'$ undergo a gradual increase with $T$ from 2 K, with a broad shoulder, say, for example at 215 K for 100 kHz, and the shoulder temperature is frequency dependent (see, for instance, Ref. 24, 25). The feature is spread over a range of ~160 – 240 K. Corresponding feature, appearing in the form of a distinct peak, is clearer in the tan$\delta$ data; one can see a systematic variation of this peak temperature with frequency, and this dispersion indicates relaxor-type ferroelectricity in this compound. We have fitted the value of the peak temperature of tan$\delta$ with Vogel-Fulcher (V-F) equation, $\nu = \nu_0 \exp[-E_a/k_B(T_m-T_g)]$, where $E_a$ is the activation energy, $k_B$ is the Boltzmann constant, $\nu_0$ is the pre-exponential factor, $T_m$ is the temperature at which tan$\delta$ peaks and $T_g$ is the freezing temperature. This fitting (see Fig. 2c) resulted in the value of $E_a$= 0.013 eV, $\nu_0$ = 1.8 x 10$^6$ Hz and $T_g$= 148 K, which agrees well with the literature[21,22]. We would like to mention here that, in the temperature range of above interest, the tan$\delta$ values are in the range of 0.001-0.01, which establishes highly insulating behaviour and negligible influence of extrinsic contributions.

Dielectric constant and the loss factor (tan$\delta$) measured with 70 kHz are shown below 300 K in Fig. 3a and Fig. 3b respectively in the presence of various magnetic fields (0, 10, 30, 50 and 70 kOe). It is clear from this figure that there is a systematic decrease in the dielectric constant as the magnetic field is increased. However, interestingly, there is no observable change in the tan$\delta$ values (see Fig. 3b) with the application of $H$; low values even in high fields establish intrinsic nature of magneto-dielectric behaviour in this compound. Fig. 3c shows the MDE (derived from the curves obtained with 20 kHz) in the form of $\Delta\varepsilon'$, where $\Delta\varepsilon' = [((\varepsilon'(H) - \varepsilon'(0))/\varepsilon'(0)]$. The values of $\varepsilon'(H)$ at 2 K and 300 K are -3.15 and -2.4 % respectively. For some fixed temperatures (2 K, 225 K and 300 K), $\Delta\varepsilon'$ measured as a function of $H$ and the curves thus obtained are presented in the inset of Fig. 3c. It is interesting that intrinsic and significant MDE is observed in a broad temperature range, even at room temperature.

**Pyrocurrent behaviour.** We did detailed pyroelectric current measurements for various protocols (temperature, magnetic field, rate dependence etc.). We first poled from 300 K to 2 K with the applied electric field of +4 kV/cm. A peak is observed around 300 K for the rate of warming ($dT/dt$= ) 2 K/ min (see Fig. 4). It is found that the direction of polarization gets reversed, if one changes the direction of the poling electric field (Fig. 4, inset). We did additional measurements as discussed in ref. 23, that is, varying the poling temperature for the same electric field and then recording the corresponding pyrocurrent behavior as a function of temperature. We obtained *I* vs *T* curves by poling from 330 K, 300 K and 230 K (to 100 K) with the electric field of 4 kV/cm. We noted (see Fig. 4) that the peak appears essentially at the respective poling temperatures; the magnitude at the peak also increases with the poling temperature. Such a behavior, rarely known in the literature, has been demonstrated for relaxor ferroelectric state in a delafossite[23]. We therefore attribute such a feature to the role played by polar nanoregions[22]. We have also obtained the pyropeak for different rates of heating, 2, 3, 4 and 5 K/min at 300 K and we found that there is an observable rate dependence (see Supplementary Fig. S3). This implies that, at room temperature, thermal stimulation effect plays a role. Interestingly enough, in the presence of various magnetic fields (10, 30 and 50 kOe), the pyro peak at this temperature arising from such an extrinsic effect shifts to higher temperature side with the increase in the magnitude as shown in inset of Fig. 4.

We have also measured polarization-electric-field loop as well as positive-up-negative-down measurements at 125 K to see ferroelectric features at this temperature. (see, Supplemenraty Fig. S4).

**$^7$Li NMR.** NMR measurements were carried out on powdered sample at 27 MHz and the spectrum was obtained by Fourier transformation. A single Gaussian line was obtained at 293 K along with asymmetric quadrupole satellites (see dotted curves in Fig. 5). The quadrupolar coupling constant turns out to be about 92 kHz at 290 K. The full width at half maximum (Δν) of the spectral line is found to be 22 kHz at 290 K. We found that the Knight shift is negligible down to 20 K. Many lithium based systems show a similar behaviour due to weak coupling with the neighbouring nuclei. However, the line-width Δν is found to vary drastically with temperature (Fig. 6) and it has a close resemblance to the variation of magnetic susceptibility. This is a strong indication that the $^7$Li NMR line-width and macroscopic magnetization have a common origin. It is notable that, below 20 K, there is a marked asymmetry in the spectra with respect to Gaussian line attributable to resonances with different widths from different Li sites (experiencing different hyperfine interactions depending on local magnetic environment, as the magnetically ordered regime is approached). Strictly speaking, in the absence of crystallographic disorder, there is only one site for Li and therefore this asymmetry supports the conclusion on the presence of anti-site Li ions in this compound[7]. The magnetic Ni ion has two sites forming distorted honeycomb lattice and linear chains with different types of magnetic interactions as revealed by neutron diffraction and some fraction of Li occupies both these sites as briefed in the introduction. The spectra well below 8 K is broadened considerably due to the transferred hyperfine field, signifying the onset of long range magnetic order. Spin-lattice relaxation time ($T_1$) behavior also is consistent with multiple Li sites. At least two distinct components, one with a $T_1$ of 10 ms and another with a $T_1$ of 3.5 ms, could be resolved at 290 K. As the temperature is decreased both the components remain distinct, and the magnitude of $T_1$ increases monotonously to 16.3 ms and 6.3 ms at 24 K. (Fig. 6, inset). The ratio of these two components remains at about 75:25 throughout the temperature range of measurement. In short, NMR thus provides evidence for crystallographic disorder, which must be the source of nanopolar behavior.

**Raman spectroscopy.** The Raman spectrum at room temperature is shown in the top panel of Fig. 7. The spectral features at 332cm$^{-1}$ and 365 cm$^{-1}$ correspond to the bending modes of the $MO_4$ tetrahedra while those at 795cm$^{-1}$, 822cm$^{-1}$, 890cm$^{-1}$, 933 cm$^{-1}$ and 962 cm$^{-1}$ correspond to the stretching modes of the $MO_4$ tetrahedra. These modes are concomitant to the bending modes (326cm$^{-1}$ and 370cm$^{-1}$) and stretching bands (823 cm$^{-1}$, 853 cm$^{-1}$, 910 cm$^{-1}$, 974cm$^{-1}$, 995cm$^{-1}$) of the isomorphic structure $Li_2Mg_2Mo_3O_{12}$ [ref. 26]. The entire spectrum was fitted with this assignment of the peaks. The temperature dependencies of the peak position of strongest spectral features $v_1$ ($365cm^{-1}$),$v_2$ ($890cm^{-1}$), $v_3$ ($933cm^{-1}$), $v_4$ ($962cm^{-1}$) are also shown in Fig. 7. All of these spectral features vary non-monotonically in the neighborhood of 180 K and 10 K as demonstrated in the right and left panels respectively.

**Discussions.** The magnetization results establish that the title compound undergoes a complex long-range magnetic order below 8 K with a magnetic compensation point at a lower temperature. Heat-capacity data reveals another feature at a higher temperature, that is, around 225 K. The dielectric constant exhibits considerable frequency dispersion in this high temperature range. This is attributable to relaxor ferroelectric behaviour. This is a new finding as far as the property of this compound is concerned. A fitting of the peak in tanδ to Vogel-Fulcher relation yields parameters which are consistent with those for relaxor ferroelectrics.

The source of this phenomenon can be attributed to nanopolar regions created by crystallographic disorder intrinsic to this structure. It is known[7] that about 25% of Li goes to Ni1 site (honeycomb site), while 50% of Li goes to Ni2 site (chain site). The present $^7$Li NMR results provide a microscopic evidence for such a disorder, as evidenced by asymmetric spectra at low temperatures with different Li ions experiencing different internal hyperfine fields (with the onset of magnetic order of Ni) as well as by spin-lattice relaxation data. It is also important to note that the charge states of Li, Ni and Mo are different (+1, +2 and +6 respectively). Therefore, the presence of antisite Li results in random local electric field in the compound. Needless to stress that relaxor ferroelectricity due the charge imbalance at a particular site in the material due to disorder has been known to be one of the main ingredients of the relaxor ferroelectric[21,22]. Therefore we conclude that the generation of the polar nano region due to charge imbalance at various sites is responsible for the observed relaxor ferroelectricity.

The spectral features in the Raman spectroscopy vary non-monotonically in the neighborhood of 180 K and 10 K and it is highly likely that these variations in the peak positions near 180K and 10K are triggered by electrostriction due to relaxor ferroelectric behaviour and magnetostriction due to onset of magnetic order respectively.

We have made an intriguing finding in the magnetocapacitance data. The values of tanδ in the entire temperature range of investigation is extremely small, even in the presence of external magnetic fields. This finding emphasizes that the observed magnetodielectric effect at room temperature is an intrinsic effect and not due to any extrinsic factors.

It may be remarked that the sensitivity of the peak temperature in the pyrocurrent plot to heating rate is usually attributed to 'thermally stimulated depolarization current'[27-31] due to random trapping of mobile careers. Since there appears to be the formation of nanopolar regions, the observed rate dependence may originate from 'dynamic disorder', possibly due to the diffusion of small Li ion at various sites, which can be controlled by thermal effects. Such a dynamic disorder may contribute features mimicking 'thermally stimulated depolarization current'.

In conclusion, the compound, $Li_2Ni_2Mo_3O_{12}$, characterized by unique crystallographic features (namely, coexisting honeycomb and linear chains of magnetic ions) and ordering magnetically below 7.2 K, is investigated by magnetization, heat-capacity, $^7$Li NMR, Raman spectroscopy, dielectric, magnetodielectric and pyroelectric current measurements. The results reveal a new feature in the range 160 – 240 K - that is, relaxor ferroelectric behaviour attributable to crystallographic disorder intrinsic to this compound), with the pyrocurrent peak temperature tracking the poling temperature. In addition, intrinsic magnetodielectric effect is seen over a broad temperature range including room temperature.

**Methods**

The polycrystalline $Li_2Ni_2Mo_3O_{12}$ was prepared by a conventional solid state route. The stoichiometric amounts of high purity (>99.9%) $Li_2CO_3$, NiO and $MoO_3$ were mixed together in an agate mortar, following by calcination at 650 $^0$C for 12 hours. The calcinated powder was then sintered at 700$^0$C for 144 hours with intermediate grindings. XRD pattern was obtained with Cu K$_\alpha$ (λ =1.54 Å).

Temperature dependent dc magnetization measurements were carried out with the help of a commercial SQUID magnetometer (Quantum Design, USA). Heat-capacity studies were carried out with a commercial Physical Properties Measurements System (PPMS) (Quantum Design, USA). Unlike otherwise stated, all these measurements were performed for the zero-field-cooled state (from 300 K) of the specimen. The same PPMS system was used to measure complex dielectric permittivity using an Agilent E4980A LCR meter with a home-made sample holder with several frequencies (ν= 1 to 100 kHz) and with a bias voltage of 1 V; this sample holder was also used for pyroelectric studies with Keithley 6517B electrometer by poling at 300 K with electric fields of 4.16 kV/cm.

$^7$Li NMR experiments were carried out at 27 MHz by sweeping the magnetic field. $T_1$ was measured by saturation recovery method using $\pi/2 - \tau - \pi/2$ pulse sequence. Raman spectra at different intervals of temperature below 300 K were recorded using an optical cryostat with continuous helium flow (MicrostatHe: Oxford instruments) and single stage adaptation 1800grooves/mm spectrometer (T64000: Horiba Jobin Yvon). The sample was excited with 488nm excitation wavelength from a mixed gas laser (Stabilite 2018: Spectra Physics).

**Acknowledgments**
The authors thank Dr. Kiran Singh for fruitful discussions.

**Author contributions**
S.K.U prepared the sample and characterized the same. He carried out magnetization, heat-capacity dielectric and pyrocurrent studies jointly with K.K.I. and analysed the results. P.L.P performed NMR studies and analysed the results. Smita Gohil and Shankar Ghosh performed Raman studies and analysed the results. E.V.S formulated the paper and finalized the manuscript in consultation with others for respective studies.

**Competing financial interests**
The authors declare no competing financial interests.

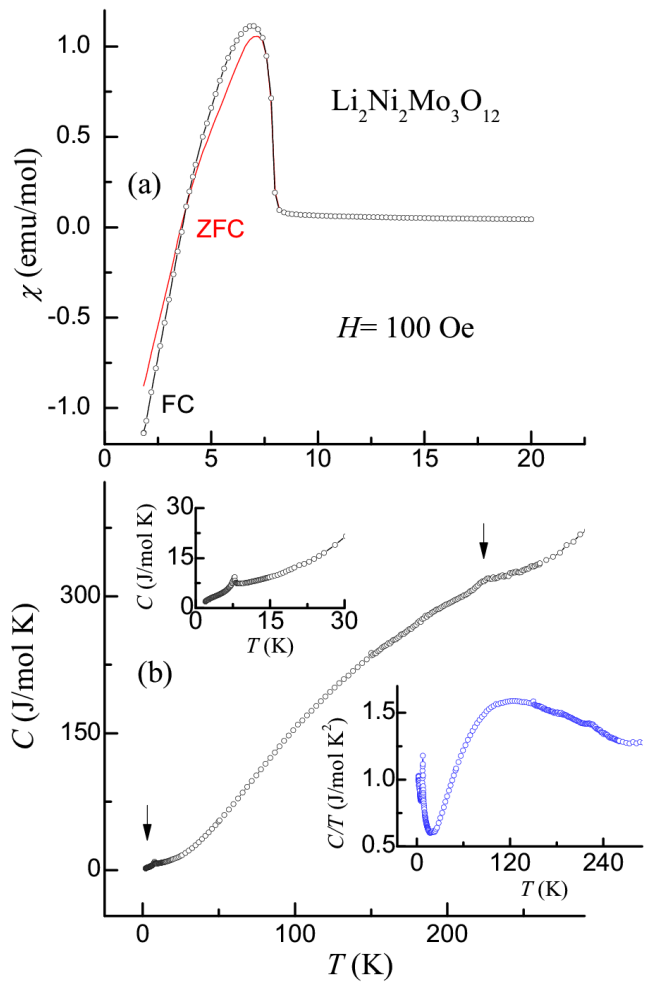

**Figure 1:** **(a)** Magnetic susceptibility obtained in 100 Oe for zero-field-cooled and field-cooled conditions of the specimen; (b) heat-capacity as a function of temperature for Li$_2$Ni$_2$Mo$_3$O$_{12}$ in the absence of magnetic field; vertical arrows mark the features of interest. Top and bottom insets in (b) show the *C/T* and *C* plots to highlight the features.

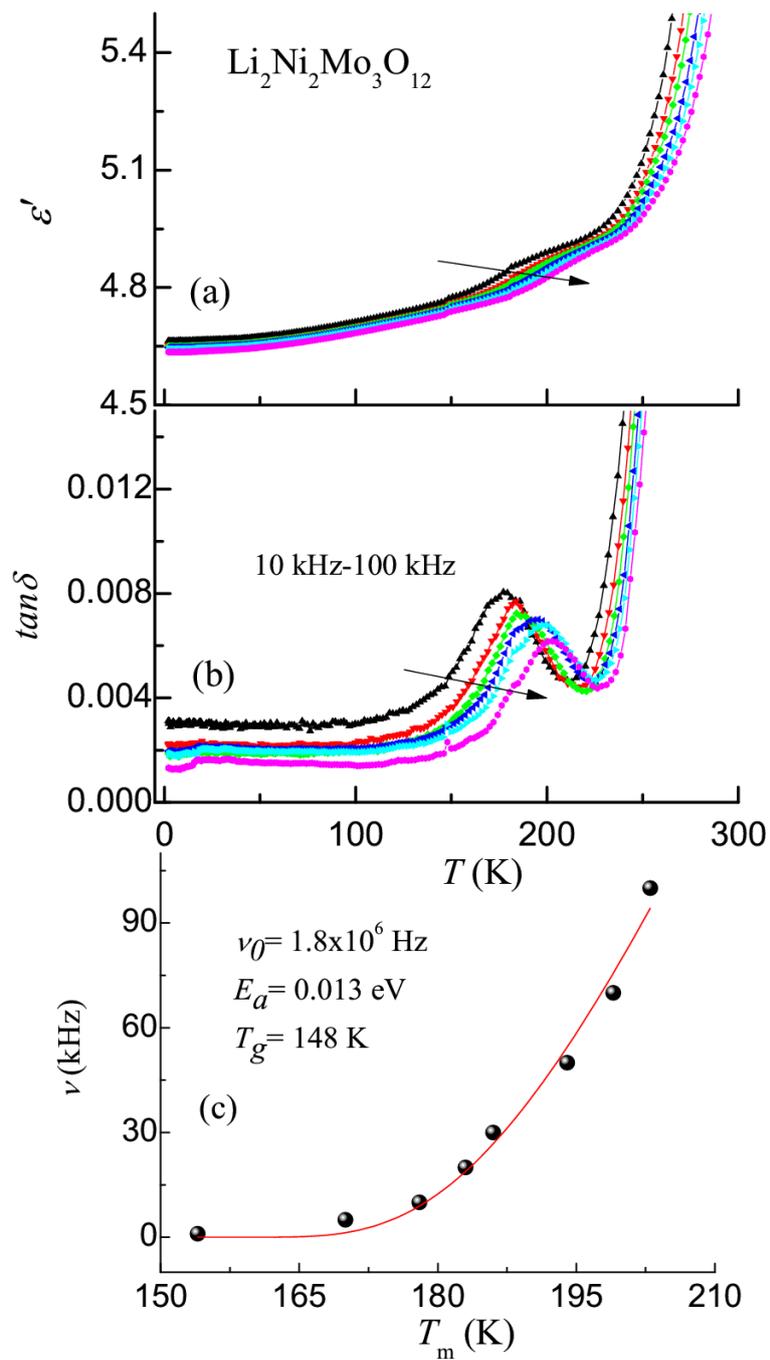

**Figure 2:** (a) Real part of the permittivity and (b) loss as a function of temperature with varying frequency for Li$_2$Ni$_2$Mo$_3$O$_{12}$. Arrows are drawn to show that the curves shift to higher temperature range as the frequency is increased (shown for 10, 20, 30, 50, 70 and 100 kHz). In (c), Vogel-Fulcher fitting for the tan$\delta$ peak values is shown.

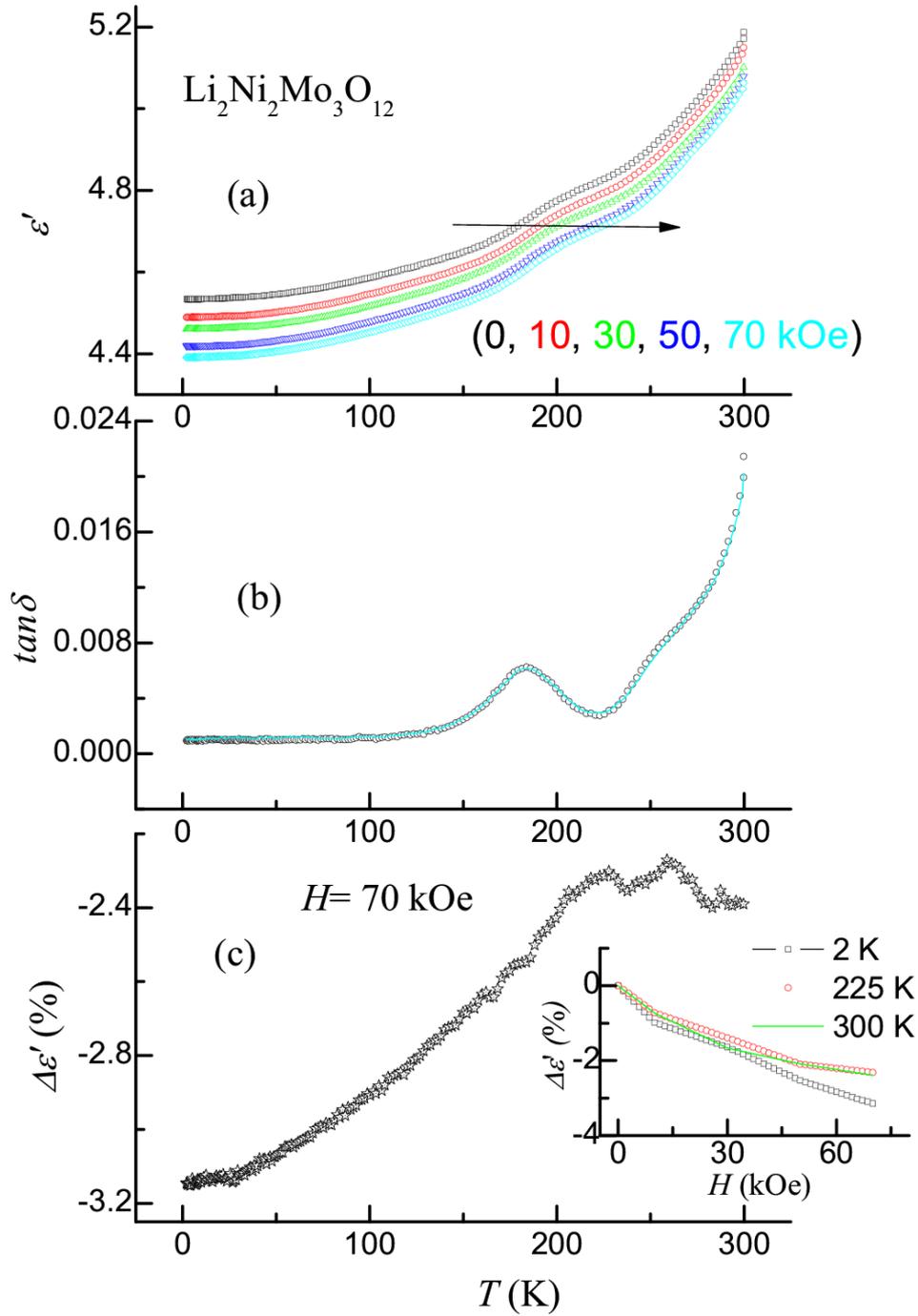

**Figure 3: (a)** Dielectric constant and (**b**) tanδ in zero field and in various magnetic fields (10, 30, 50 and 70 kOe), measured with 70 kHz, for $Li_2Ni_2Mo_3O_{12}$. In (**b**), data points correspond to zero field, while in-field behavior (70 kOe) is represented by a continuous line. Both these curves overlap. **(c)** Magnetodielectric effect (Δε′) for a change of magnetic field from zero to 70 kOe as extracted from the data shown in (**a**). Inset of (**c**) shows Δε′ versus $H$ for some selected temperatures.

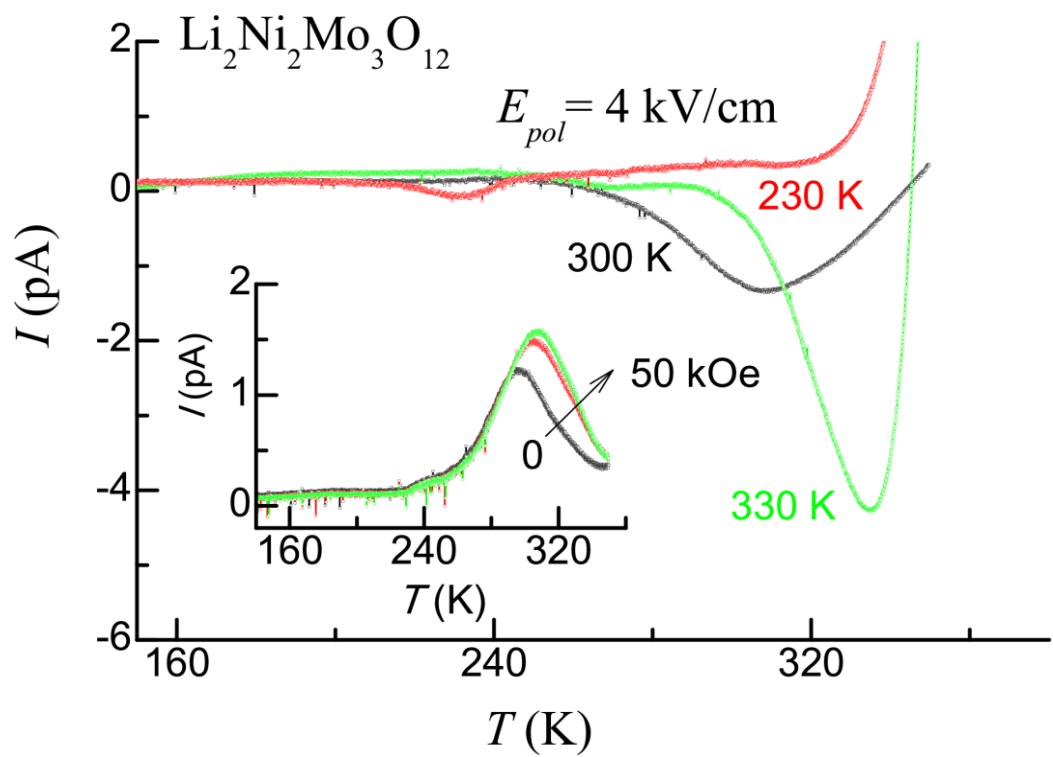

**Figure 4:** Pyrocurrent behaviour obtained with a poling field of 4kV/cm for $Li_2Ni_2Mo_3O_{12}$ for poling temperatures of 230, 300 and 330 K. Inset shows pyrocurrent behavior in the presence of magnetic fields at 300 K for the poling electric field of -4kV/cm and an arrow is drawn to show that the curves move towards higher temperature range with increasing magnetic field with the curves corresponding to 0, 30 and 50 kOe

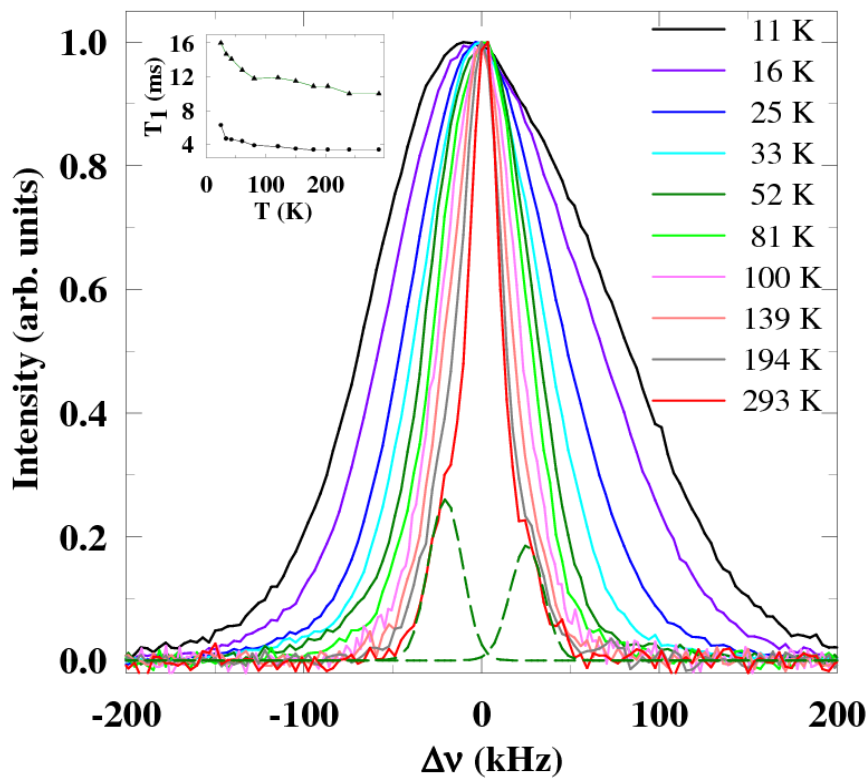

**Figure 5:** Temperature dependence of the $^7$Li NMR spectra for Li$_2$Ni$_2$Mo$_3$O$_{12}$ obtained at ($\nu$=) 27 MHz. The origin of the x-axis $\Delta\nu = 0$ corresponds to the centre of the spectra at 293 K. The dotted green lines represent the asymmetric satellites at 293 K. The inset shows the $T_1$ variation with temperature. The solid line is guide to eyes.

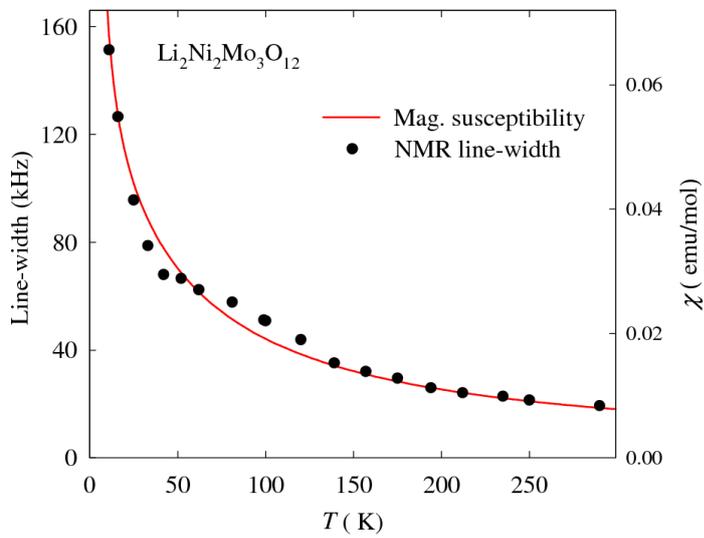

**Figure 6:** The variation of magnetic susceptibility and NMR line-width as a function of temperature for Li$_2$Ni$_2$Mo$_3$O$_{12}$.

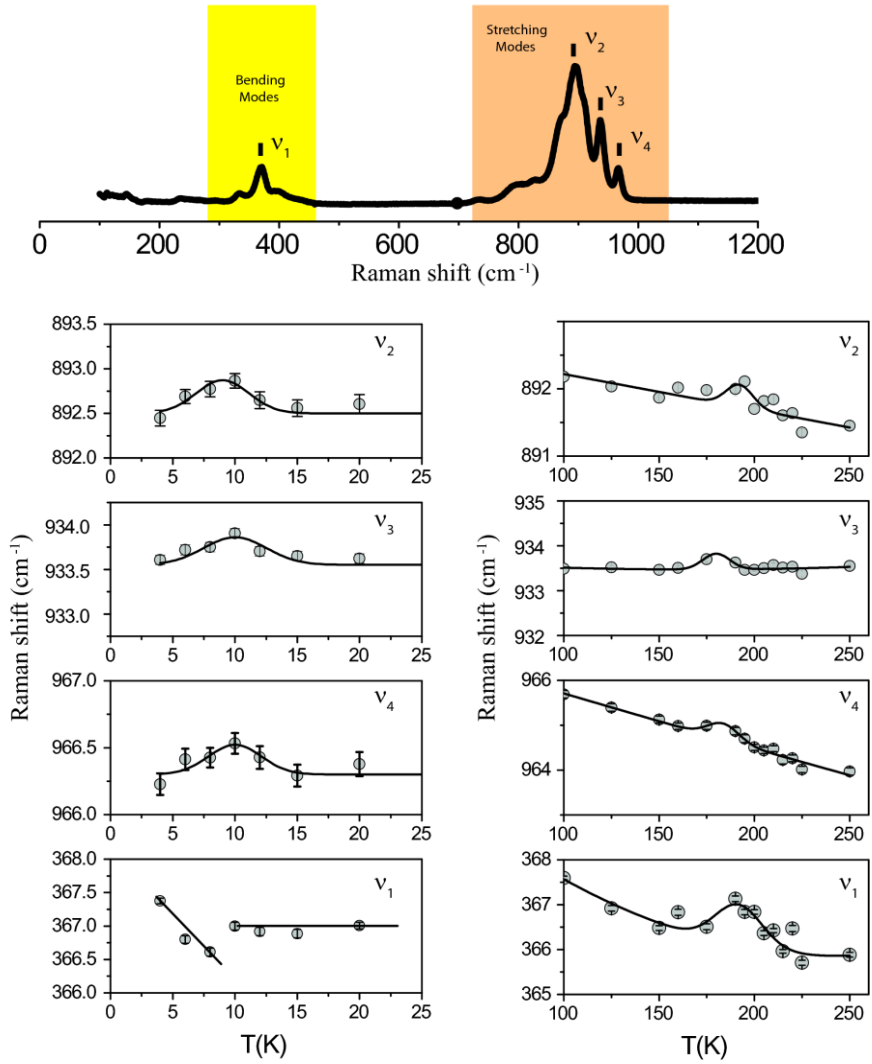

**Figure 7:** The room temperature Raman spectrum of $Li_2Ni_2Mo_3O_{12}$ obtained by exciting the sample with an incident laser beam of wavelength 488 nm is shown in the top panel. The bottom left and right panels show the variation of the spectral features corresponding to $\nu_1, \nu_2, \nu_3, \nu_4$ in the tempurature range 4.2 to 25 K and 100 to 250K. The solid lines are guides to the eyes.

# Supplementary Information
## Pyrocurrent anomalies and intrinsic magnetodielectric behavior near room temperature in Li$_2$Ni$_2$Mo$_3$O$_{12}$, a compound with distorted honeycomb and spin-chains

**Sanjay Kumar Upadhyay, Kartik K Iyer, Smita Gohil, Shankar Ghosh, P.L. Paulose, and E.V. Sampathkumaran***

Tata Institute of Fundamental Research, Homi Bhabha Road, Colaba, Mumbai 400005, India

*Corresponding author: sampath@mailhost.tifr.res.in

**X-ray diffraction data:**

In this section, we show the x-ray diffraction pattern in Fig. S1 along with Rietveld refinement and fitted parameters. We used the same structural parameters as that employed in ref. 8.

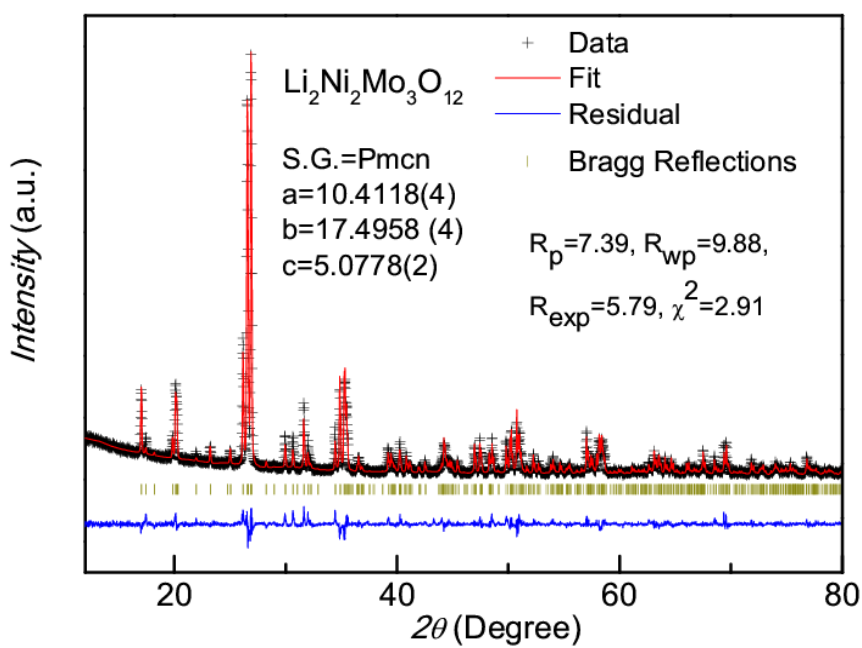

**Supplementary Figure S1:** XRD pattern (obtained at room temperature) along with the Reitveld refinement for Li$_2$Ni$_2$Mo$_3$O$_{12}$.

**Magnetization data:**

In order to show that the magnetization results obtained by us on this material are in good agreement with that reported in the literature, we show magnetic susceptibility curves and isothermal magnetization behaviour in Fig. S2. We made additional new findings, that is, the observation of a weak hysteresis around 40 kOe as well as the existence of a weak loop at low fields for 6 K.

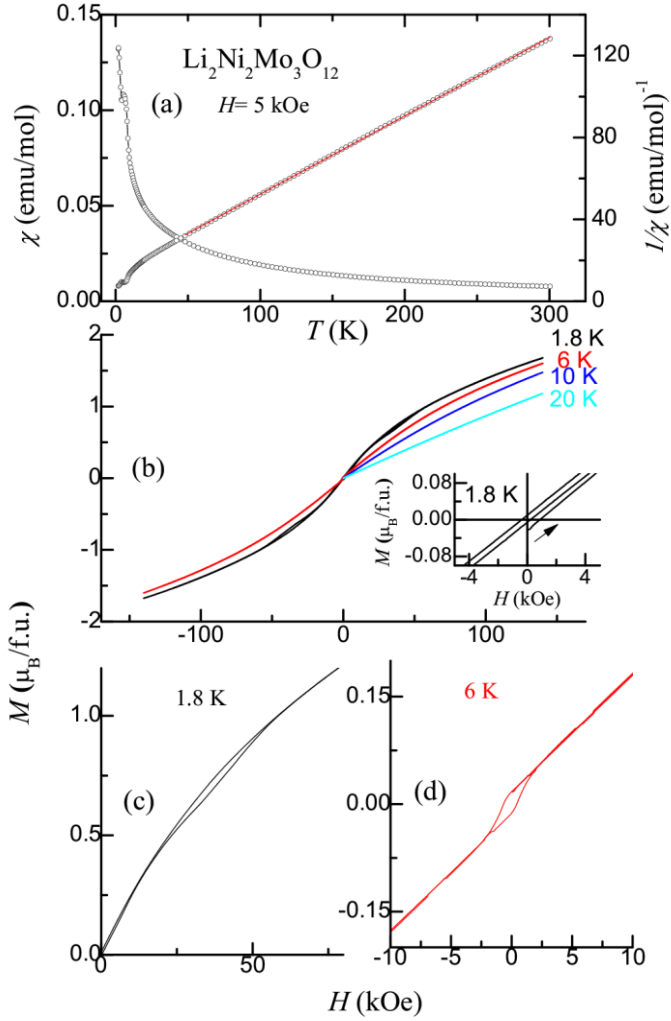

**Supplementary Figure S2:** (a) Magnetic susceptibility ($\chi$) and inverse $\chi$ obtained in the presence of 5 kOe for $Li_2Ni_2Mo_3O_{12}$. In the inverse $\chi$ plot, the line drawn is obtained by Curie-Weiss fitting. In the case of $\chi$ versus $T$ plot, a line is drawn through the data points to serve as a guide to the eyes. (b) Isothermal magnetization at selected temperatures. Top inset in (b) shows the *M(H)* at low fields at 1.8 K to highlight that the virgin curve lies outside envelope curve. (c) shows isothermal magnetization behaviour up to 70 kOe magnetic field at 1.8 K. (d) The M(H) curve in the low field region is shown in an expanded form for 6 K to highlight the existence of a low-field loop.

**Pyrocurrent behavior:**

In this section, we show the dependence of pyrocurrent peak on the rate of change of temperature. We have poled from 300 K to 2 K with the applied electric field of 4 kV/cm and collected the data for the different rates of warming ($dT/dt=$ ) 2 K/min, 3 K/min, 4 K/min and 5 K/min (see Fig. S3) from 2 K in independent measurements. We observe a peak around 300 K, but there is an observable rate dependence of the peak temperature.

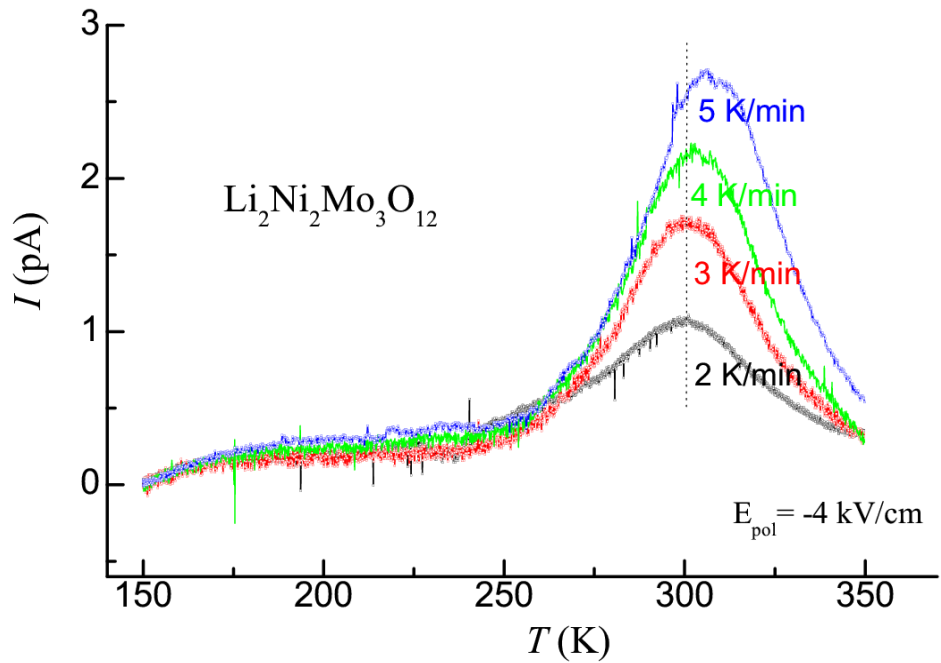

**Supplmentary Figure S3:** Pyrocurrent behaviour for $Li_2Ni_2Mo_3O_{12}$ for different heating rates. A dotted vertical line is drawn to show that the peak shifts with higher rate of heating.

**Additional information:**

The ferroelectric measurements (P-E loop and PUND) were carried out at 125 K (that is, below $T_g$) by a P-E loop tracer supplied by M/s Radiant Instruments, USA, coupled to a home-made sample holder. The PE loop is shown in Fig. S4. Due to the difficulties in applying adequate high electric fields to enable us to see the saturation of the loop, we could only apply a maximum electric field of 10 kV/cm. We see a non-linear type of PE loop, similar to that seen in other relaxor ferroelectrics (see, for instance, refs. 24-25). Usually, other than the switched charge density, non-ferroelectric factors (leakage current, free charge carrier etc.) can also contribute to PE loop. Therefore, in order to look for intrinsic ferroelectric behaviour, 'positive-up-negative-down (PUND)' measurements[S1] were performed, by applying pulses of electric field of certain pulse-width and delay (see, for instance, ref. Ref. S2). Electric polarization is measured as a function of time during this process. Intrinsic polarization due to ferroelectricity is given by $\Delta P = P1-P2$, also defined in the inset of Fig. S4. This turns out to be finite and hence this data could support ferroelectric nature of this material at such temperatures.

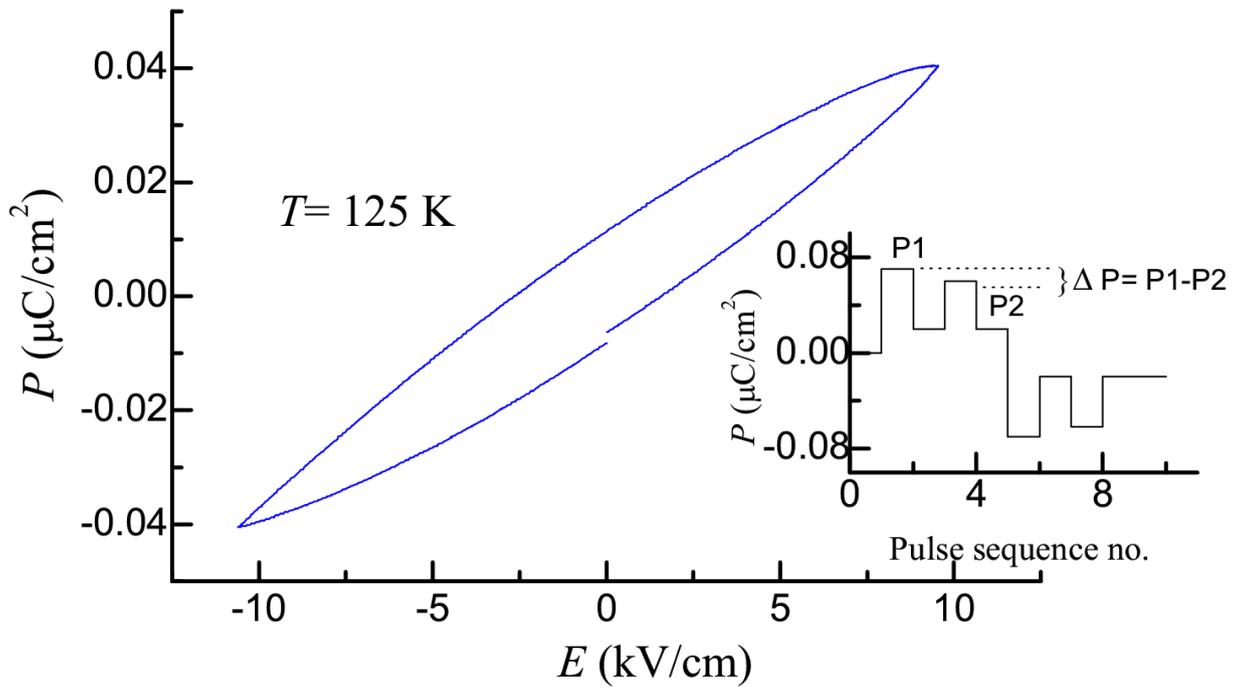

**Supplementary Figure S4:** PE loop at 125 K obtained with 50 Hz. Inset shows the PUND plot (electric polarization versus electric-field pulse number), obtained with a pulse width of 20 ms and delay time of 10 ms.